\documentclass[aps,pra,twocolumn,showpacs,amsmath]{revtex4}
\usepackage{bm}
\usepackage{graphicx}
\begin{document}

\title{Decoherence of a two-state atom driven by coherent light}
\author{Hyunchul Nha and H. J. Carmichael}
\affiliation{Department of Physics, University of Auckland, Private Bag 92019,
Auckland, New Zealand} 

\begin{abstract}
Recent studies of the decoherence induced by the quantum nature of the laser field
driving a two-state atom [J. Gea-Banacloche, \pra {\bf 65}, 022308 (2002); S. J. van
Enk and H. J. Kimble, Quantum Inf.\ and Comp.\ {\bf 2}, 1 (2002)] have been questioned
by Itano [W. M. Itano, \pra {\bf 68}, 046301 (2003)] and the proposal made that all
decoherence is due to spontaneous emission. We analyze the problem within the formalism
of cascaded open quantum systems. Our conclusions agree with the Itano proposal. We show
that the decoherence, nevertheless, may be divided into two parts---that due to forwards
scattering and to scattering out of the laser mode. Previous authors attribute the
former to the quantum nature of the laser field.
\end{abstract}
\maketitle

\narrowtext

\section{Introduction}
\label{sec:introduction}
Quantum computation, which relies on principles of quantum mechanics that find no
analogue in a classical computer, has attracted much attention in recent years, and
considerable effort has been devoted to implementing the elements of a quantum computer.
Some quantum algorithms and quantum logic gates have been experimentally realized;
for example, in an NMR system \cite{Vandersypen} and with trapped ions
\cite{Gulde,Schmidt,Leibfried}. Trapped-ion systems, in particular, are considered
promising, because of their scalability. The use of laser fields to manipulate
two-state atomic transitions is crucial to these systems, since such manipulations
realize the one- and two-qubit operations that form the basic building blocks
of a quantum computation \cite{Cirac}. When modeling these manipulations, the laser
fields are mostly taken to be classical; thus it is implicitly assumed that the
interaction of the atom with the laser field contributes nothing to its
decoherence. The same assumption, in fact, prevails beyond the quantum computation
context, wherever the manipulation of an atomic state by a laser field is involved. 
 
At a fundamental level, the laser should be treated as a quantum field, in which case
its interaction with the atom would generally lead to entanglement of the two. It is
natural then to ask whether the resolution of this entanglement (tracing out the laser
field) leads to additional decoherence, i.e., in addition to the decoherence due to
spontaneous emission. The issue can be framed from a quantum measurement point of view.
One may measure the sum of the laser field and forwards-scattered field from the atom.
This measurement would give, if not complete, then some information on the atomic state.
The available information, whether the measurement is actually performed or not, would
degrade the coherence of the atomic state, much as which-way information does in the
Young's two-slit experiment. We are lead, therefore, to the question posed. The issue
becomes particularly pressing when one considers a case where the atoms (ions) of a
quantum computer are to be individually addressed by a tightly focused laser beam
\cite{Schmidt,Cirac}. The solid angle subtended by the laser field is then substantial,
and a significant amount of information about the atomic qubits might be retrieved by
measuring the forwards-scattered laser light.

The question raised has been asked by previous authors, with some disagreement in the
answers \cite{Julio,Enk,Itano,Julio1,Enk1}. In this paper we address it in an alternative
analysis, in a way that, for us, leads quickly to a clear conclusion, and helps resolve
any remaining disagreement. We use the theory of cascaded open systems
\cite{Kolobov,Car,Gardiner}, first in its quantum trajectory formulation \cite{Car}, and
then as the starting point to derive a master equation for the driven atom. We show from
the derived master equation that the rate of atomic decoherence is, indeed, given entirely
by the rate of spontaneous emission for an atom driven by a classical field. We use quantum
trajectories to expose the different scattering processes that contribute to the total
rate: forwards scattering---scattering that overlaps the propagating laser pulse---and side
scattering out of the laser pulse. Our treatment of the laser pulse is one dimensional,
and therefore idealized. It nevertheless captures the essential character of the physics
involved---that the driving of an atom by a laser field is a {\it scattering process\/}.
Decoherence accounts for the resolution of the Schr\"odinger entanglement between the
scattering center and scattered fields. Assuming the driving field to be coherent, the
latter, in simple language, is due to spontaneous emission. Our approach is applicable
to nonclassical driving fields as well, where additional entanglement does arise.

We review the background in Sec.~\ref{sec:background} and formulate our treatment of
the problem in Sec.~\ref{sec:cascaded_systems}. Analysis of the developed model is
carried out in Secs.~\ref{sec:quantum_trajectories} and \ref{sec:master_equation}.
Discussion of our results and their connection with earlier work is presented in
Sec.~\ref{sec:discussion}.
 
\section{Background}
\label{sec:background}
As mentioned, several authors have considered the entanglement of a coherently driven
atom with its driving field \cite{Julio,Enk}. The previous works are carried out within
the framework of the Jaynes-Cummings model. Their main result is that the entanglement
is nonzero, although small for strong fields, scaling as $1/\bar n$, where $\bar n$ is
the mean number of photons in the laser pulse. The result has been questioned by Itano
\cite{Itano}, who argues that there is no entanglement with the driving field at all.
His argument is based on Mollow's treatment of resonance fluorescence \cite{Mollow},
which, considering a quantized driving field in a coherent state, applies a displacement
operator to the field to show that the atom equivalently couples to a classical driving
field plus the quantized vacuum. Itano suggests that all entanglement originates in
the interaction with the vacuum and is therefore resolved through spontaneous emission.

The first and single most important question arising from Itano's comment is whether
or not the entanglement reported in \cite{Julio,Enk} results in decoherence that is
additional to, or included within, the decoherence rate obtained from the standard
treatment of spontaneous emission for an atom driven by a classical field. The assertion
(implication) of Refs.~\cite{Julio,Enk} is that it is additional to; Gea-Banacloche
states so explicitly \cite{Julio1}: ``$\ldots$ I wanted to focus, instead, on the
decoherence due to the quantum nature of the laser field, which I took to be a
{\it separate\/} source of error.'' Itano's position is that there is no decoherence
in addition to spontaneous emission.

There is a second, more subtle question, featured most clearly in the reply of
van Enk and Kimble \cite{Enk1}. Does the driven atom become entangled with the laser
field at all? Considering, for sake of argument, that the total decoherence rate can be
calculated as Itano claims, can any part of it be attributed to entanglement between
the laser field and atom? By implication, if not directly, Itano claims such
entanglement is zero. The authors of the criticized work claim it is not, though
the entanglement is small \cite{Julio1,Enk1}, agreeing at most that the spontaneous
emission calculation gives the correct number for the total decoherence rate, while
asserting that its account of the laser-entanglement part of the decoherence is
incorrect.

We return to these questions in Sec.~\ref{sec:discussion}, after presenting our own
analysis of the problem of a two-state atom driven by coherent light.

Itano's criticism of Refs.~\cite{Julio,Enk} begins with the observation that the
Jaynes-Cummings model is inappropriate for treating a free-space atom driven by
a quantized field. Potential problems with the model were discussed by Silberfarb
and Deutsch \cite{Silberfarb}. In response it has been pointed out that a careful
use of the model can nevertheless yield correct results \cite{Julio1,Enk1}. We do
not intend to focus on the limitations of the Jaynes-Cummings model; we think them
amply clear. The fundamental character of the considered physics is that of a
scattering process. The Jaynes-Cummings model does not deal with scattering and
is inappropriate in this sense, though it can give meaningful results of a
perturbative sort. Our plan is to start with a formulation in explicit scattering
terms. For this purpose we model the atom and its driving field as cascaded open
quantum systems (COQS) \cite{Kolobov,Car,Gardiner}. Our model describes the
unidirectional coupling of the laser output field and a target two-state atom in
the Born-Markov approximation. The quantized field interacting with the atom is
intrinsically multimode; it supports a propagating laser pulse and an outgoing
scattered field. The Lindblad master equation derived by Silberfarb and Deutsch
\cite{Silberfarb} is contained within the model through its forwards scattering
terms. While we consider only coherent driving of the atom here, the COQS approach
is more general and can treat nonclassical driving fields as well.

\section{Cascaded Quantum Systems}
\label{sec:cascaded_systems}
\begin{figure}[b]
\includegraphics*[width=2.75in,keepaspectratio=true]{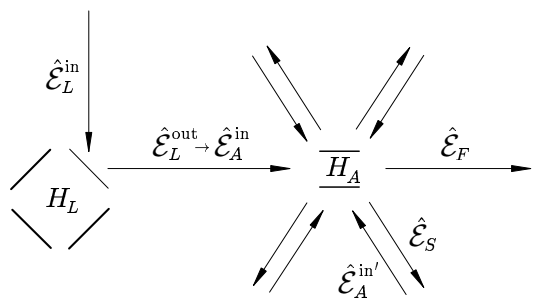}
\caption{Schematic diagram of the cascaded system of a two-state atom (Hamiltonian
$H_A$) driven by a coherent laser source (Hamiltonian $H_L$). The various inputs
and outputs are defined in the text.}
\label{fig:fig1}
\end{figure}
We begin by briefly reviewing the COQS model. As depicted in Fig.~\ref{fig:fig1}, the
complete system is comprised of the laser subsystem, denoted $L$, and the target atom
subsystem, denoted $A$. The subsystems have free Hamiltonians $H_L$ and $H_A$,
and couple through the quantized electromagnetic field, which is denoted as a reservoir
with Hamiltonian $H_R$. Free-fields $\hat{\cal E}_L^{\rm in}$ and $\hat{\cal E}_A
^{{\rm in}}{}^\prime$ provide vacuum inputs to subsystems $L$ and $A$, respectively.
Subsystems $L$ and $A$ couple unidirectionally through the common channel $\hat{\cal E}_L
^{\rm in}\to\hat{\cal E}_L^{\rm out}\to\hat{\cal E}_A^{\rm in}\to\hat{\cal E}_F$. The
scattered fields are the forwards scattered field $\hat{\cal E}_F$ and the sideways
scattered field $\hat{\cal E}_S$. All fields have units of the square root of photon
flux.

A master equation for $L\otimes A$ is derived in the Born-Markov approximation
\cite{Car}. First, the laser output field in the Heisenberg picture is written as
\begin{equation}
\hat{\cal E}_L^{\rm out}(t)=\hat{\cal E}_L^{\rm in}(t)+\sqrt{2\kappa_L}\mkern2mu
\hat a(t),
\label{eqn:output}
\end{equation} 
where $\hat a(t)$ is the annihilation operator for the intracavity mode and
$2\kappa_L$ is the cavity linewidth. This field propagates freely to the
target atom, which thus couples to the retarded field $\hat{\cal E}_L^{\rm out}(t')$,
$t'=t-l/c$, where $l$ is the distance from $L$ to $A$. Designating the position of the
output mirror as $z=0$, the field $\hat{\cal E}_L^{\rm out}(t)$ is evaluated at $z=0+$.
The laser cavity mode couples to the field at $z=0$. Denoting the latter by
$\hat{\cal E}_L(t)$, the atom thus couples to the field
\begin{equation}
\hat{\cal E}_L^{\rm out}(t^\prime)=\hat{\cal E}_L(t^\prime)+{\textstyle\frac12
\displaystyle}\sqrt{2\kappa_L}\mkern2mu\hat a(t^\prime).
\end{equation}
We note that the atom and laser cavity mode (at the retarded time) both couple to the
reservoir field $\hat{\cal E}_L(t^\prime)$; the atom also couples directly to the intracavity
laser field $\frac12\sqrt{2\kappa_L}\mkern2mu\hat a(t^\prime)$ and to the reservoir
field $\hat{\cal E}_A^{\rm in}{}^\prime$.

Let us now return to the Schr\"odinger picture and denote the density operator for the
entire system, subsystems $L$ and $A$ plus the reservoir $R$, by $\chi(t)$, and define
the source-retarded density operator
\begin{equation}
\chi'(t)\equiv U_L(l/c)\chi(t)U_L^{\dag}(l/c),
\end{equation}
with
\begin{equation}
U_L(l/c)\equiv \exp[(i/\hbar)(H_L+H_R+H_{LR})(l/c)],
\end{equation}
where $H_{LR}$ is the interaction between $L$ and $R$. Then from the reservoir couplings
noted above, applied formally at $t^\prime=0$, by a standard derivation, the equation of
motion for the reduced density operator $\rho'\equiv{\rm tr}_R(\chi')$ is \cite{Kolobov,Car,Gardiner} 
\begin{eqnarray}
{\dot \rho^\prime}=\frac{1}{i\hbar}[H_0,\rho^\prime]+{\cal L}_{\hat J_F}\rho^\prime
+{\cal L}_{\hat J_S}\rho^\prime,
\label{eqn:master}
\end{eqnarray}
where ${\cal L}_{\hat O}$ is the Lindblad superoperator,
\begin{eqnarray}
{\cal L}_{\hat O}\equiv\hat O\cdot \hat O^{\dag}-\frac12\hat O^{\dag}\hat O\cdot
-\frac12\cdot \hat O^{\dag}\hat O,
\label{eqn:superoperator}
\end{eqnarray}
and
\begin{equation}
H_0=H_L+H_A+i\hbar\sqrt{\kappa_L\kappa_A}\left(\hat a^{\dag}\hat\sigma_-
-\hat a\hat\sigma_+\right),
\label{eqn:Hamiltonian}
\end{equation}
where $\hat\sigma_+$ and
$\hat\sigma_-$ are raising and lowering operators for the atom. The forwards- and
side-scattering jump operators are
\begin{equation}
{\hat J}_F=\sqrt{2\kappa_L}\mkern2mu\hat a+\sqrt{2\kappa_A}\mkern2mu\hat\sigma_-,\qquad
{\hat J}_S=\sqrt{2\kappa^\prime_A}\mkern2mu\hat\sigma_-.
\label{eqn:jump}
\end{equation}

The Lindblad ${\cal L}_{\hat J_F}$ enters Eq.~(\ref{eqn:master}) through the coupling
of  $L$ and $A$ to the common (forwards) scattering channel via the field
${\cal E}_L(t^\prime)$, where a decay rate $2\kappa_A$ is introduced phenomenologically
to parameterize the strength of the coupling to the atom; this coupling strength depends
on the overlap of the atomic dipole mode with the solid angle subtended by the laser
pulse. The Lindblad ${\cal L_{\hat J_S}}$ accounts for the interaction of the atom with
the additional reservoir field $\hat{\cal E}_A^{\rm in}{}^\prime$, and contributes the
additional decay rate $2\kappa^\prime_A$. The total free-space atomic decay rate is
$\gamma=2\kappa_A+2\kappa^\prime_A$.

Equations (\ref{eqn:master})-(\ref{eqn:jump}) define the model we adopt as the
simplest implementation of the driving of a two-state atom by a quantized field as a
scattering process. The model introduces a number of approximations. In particular, the
laser field is treated as one-dimensional and single mode, both potentially severe
restrictions for a highly focused laser pulse. The model is adequate, however, to address
the questions raised, and its generalization is straightforward. Relaxation of the
one-dimensional approximation, for example,  is discussed in \cite{Enk2}.

\section{Quantum Trajectories}
\label{sec:quantum_trajectories}
The scattering features of our model are most apparent in a quantum trajectory unraveling
of master equation~(\ref{eqn:master}) \cite{Car,Carmichael93}. One imagines performing
photoelectric detection of the fields $\hat{\cal E}_F$ and $\hat{\cal E}_S$ and making a
record, ${\rm REC}$, of the times and type of the detection events. One constructs a
conditional evolution, conditioned on the detection record, in which the (unnormalized)
stochastic wavefunction $|\bar\psi_{\rm REC}\rangle$ evolves continuously according to
the Schr\"odinger-like equation
\begin{equation}
|\dot{\bar{\psi}}_{\rm REC}\rangle=\frac1{i\hbar}H_B|\bar\psi_{\rm REC}\rangle,
\label{eqn:schroedinger}
\end{equation}
with non-Hermitian Hamiltonian
\begin{eqnarray}
H_B&=&H_L+H_A-i\hbar\kappa_L\hat a^{\dag}\hat a
-i\hbar\frac\gamma2\hat\sigma_+\hat\sigma_-\nonumber\\
&&-2i\hbar\sqrt{\kappa_L\kappa_A}\mkern2mu\hat a\hat\sigma_+,
\label{eqn:sde}
\end{eqnarray}  
and suffers quantum jumps,
\begin{subequations}
\begin{equation}
|\bar\psi_{\rm REC}\rangle\rightarrow\hat J_F|\bar\psi_{\rm REC}\rangle,
\end{equation}
\begin{equation}
|\bar\psi_{\rm REC}\rangle\rightarrow\hat J_S|\bar\psi_{\rm REC}\rangle,
\label{eqn:jump_side}
\end{equation}
\end{subequations} 
determined in a Monte-Carlo fashion with probabilities (per time step $dt$)
\begin{subequations}
\begin{equation}
p_{LA}=\langle\psi_{\rm REC}|\hat J_F^{\dag}\hat J_F|\psi_{\rm REC}\rangle dt,
\end{equation}
\begin{equation}
p_A=\langle\psi_{\rm REC}|\hat J_S^{\dag}\hat J_S|\psi_{\rm REC}\rangle dt.
\end{equation}
\end{subequations}
Jumps executed by the operators $\hat J_F$ and $\hat J_S$ denote the detection
of a photon in the forwards- and side-scattered fields, respectively ($\hat{\cal E}_F$
and $\hat{\cal E}_S$ in Fig.~\ref{fig:fig1}).

In this formulation the unidirectional coupling of the laser source to the target atom
is explicit. The last term, proportional to $\hat a\hat\sigma_+$, in
Hamiltonian~(\ref{eqn:sde}) annihilates a laser photon and excites the atom, but there
is no term $\hat a^\dagger\hat\sigma_-$ for the reemission of photons into the laser
field. Thus, as Itano pointed out, interaction with the atom does not change the laser
field {\it upstream\/} from the atom (the time-retarded field at the source). The atom
does, however, absorb photons emitted a retardation time earlier by the laser source.
Photons are annihilated {\it downstream\/} from the atom (at the imagined detector) by
the jump operator $\hat J_F$. It is not possible to distinguish between the straight-through
laser field and forwards reemission, and this indistinguishability can, in principle,
give rise to entanglement. For example, for a Fock state $|n\rangle_L$ of the laser mode,
under the jump $\hat J_F$ we have
\begin{equation}
|n\rangle_L|+\rangle_A\to\sqrt{2\kappa_Ln}\mkern2mu|n-1\rangle_L|+\rangle_A+
\sqrt{2\kappa_A}\mkern2mu|n\rangle_L|-\rangle_A,
\label{eqn:Fock}
\end{equation}
where $|+\rangle_A$ and $|-\rangle_A$ denote the upper and lower atomic states. In this
regard a laser pulse in a coherent state is special, since for a coherent state, as an
eigenstate of the annihilation operator $\hat a$, there is no entanglement produced by
$\hat J_F$; neither does the continuous evolution (\ref{eqn:schroedinger}) produce
entanglement, due to the absence of a term $\hat a^\dagger\hat\sigma_-$ in $H_B$. Thus,
we see that under coherent excitation the atom does not entangle with the field that
drives it (the retarded field at the source), though for excitation by a nonclassical
field it generally would. The question of whether there is entanglement with the forwards
scattering remains. We return to this question in Sec.~\ref{sec:discussion}.

We move now to the question of the decoherence rate. The side-scattering quantum jump
(\ref{eqn:jump_side}) shows how decoherence works in quantum trajectory theory for
spontaneous emission. The atom jumps to its ground state at a determined rate, which
destroys induced coherence. The rate is proportional in our model to $2\kappa_A^\prime$,
rather than the free-space spontaneous emission rate $\gamma$; the missing $2\kappa_A$
comes from the forwards scattering channel. Decoherence for the forwards scattering
operates differently, since the jump operator $\hat J_F$ does not put the atom in its
ground state. In fact, the unique form of the $\hat J_F$ jump is of interest for a
separate reason. In the absence of a term $\hat a^\dagger\hat\sigma_-$ in the Hamiltonian
$H_B$, we might wonder how there can be any Rabi oscillation? Rabi oscillations are
produced by absorption from {\it and reemission into\/} the driving field; the
reemission term is missing.

\begin{figure}[t]
\includegraphics*[width=2.75in,keepaspectratio=true]{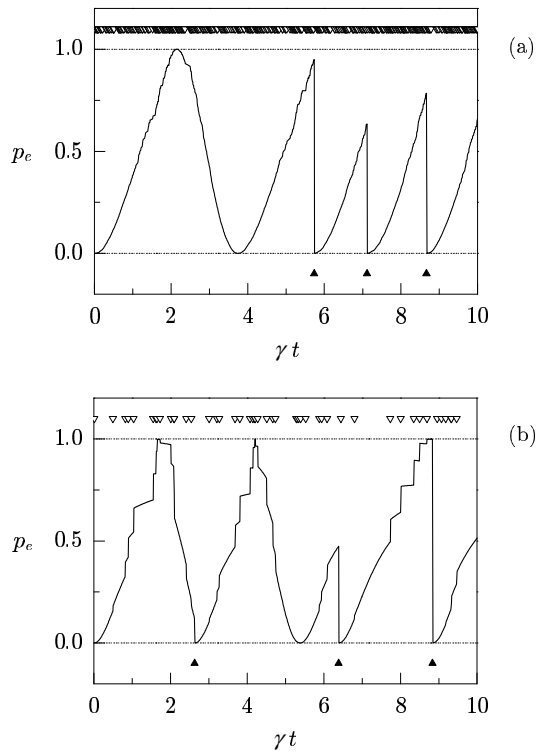}
\caption{Sample Monte-Carlo simulations of the Rabi oscillation of the driven
two-state atom in the presence of spontaneous emission, total emission rate
$\gamma=2\kappa_A+2\kappa_A^\prime$: for a Rabi frequency
$\Omega=\sqrt{\kappa_L\kappa_A}\mkern2mu|\alpha|=2\gamma$ (for coherent state
$|\alpha\rangle_L$) and forwards scattering rate (a) $2\kappa_A/\gamma=0.04$,
(b) $2\kappa_A/\gamma=0.4$. The probability $p_e(t)$ to find the atom in the excited
state is plotted a function of time. Open (closed) triangles mark the times of
forwards- (side-) scattering quantum jumps.}
\label{fig:fig2}
\end{figure}

The answer to this conundrum is that reemission, the deexitation of the atom, is
generated, not by any Hamiltonian, but by repeated $\hat J_F$ jumps \cite{Car}; it is 
irreversible, as one might expect for a scattering process. Importantly, for fixed
Rabi frequency, the form of the oscillation depends on the ratio $2\kappa_A/\gamma$,
i.e., on the degree of focusing of the laser pulse. A weakly (strongly) focused pulse
must contain more (fewer) photons to produce the same Rabi frequency. Consequently,
the inference about the sate of the atom on detecting a photon in the forwards
direction is weaker (stronger), while the frequency of these detections is higher
(lower). The inferred conditional evolution of the atomic state then depends on
$2\kappa_A/\gamma$ in the manner shown in Fig.~\ref{fig:fig2}. A weakly focused pulse
containing many photons yields an almost continuous Rabi oscillation of the kind
produced by the Jaynes-Cummings Hamiltonian with classical driving field [frame (a)].
Strong focusing and fewer photons yields a ragged oscillation since a stronger
inference about the atomic state can be made on the basis of a single photoelectric
detection [frame (b)].

It is clear from this discussion that decoherence for a coherently driven two-state
atom can be divided into two distinct parts---that due to forwards scattering and
that due to scattering out of the forwards channel. The forwards part, moreover, is
subject to a $1/\bar n$ effect as the previous authors have claimed
\cite{Julio,Enk,Julio1,Enk1}. 
 
\section{Master Equation for the Atom}
\label{sec:master_equation}
Considering the different forms of the quantum jumps in the two channels, it is
unclear whether the total decoherence rate may be considered to be due to spontaneous
emission alone or not. To resolve this issue we set the quantum trajectory
formulation aside and derive a master equation for the atom alone.

We note first, from Eq.~(\ref{eqn:master}), that the equation of motion for the
reduced density operator of $L$, $\rho_L={\rm tr}_A(\rho')$, is given by the laser
master equation
\begin{eqnarray}
{\dot \rho_L}=\frac{1}{i\hbar}[H_L,\rho_L]+{\cal L}_{\hat J_L}\rho_L,
\label{eqn:master3}
\end{eqnarray}
with
\begin{equation}
{\hat J}_L=\sqrt{2\kappa_A}\mkern2mu\hat a.
\end{equation} 
To generate a coherent state of the intracavity field, we adopt the model Hamiltonian 
\begin{eqnarray}
H_L=\hbar\omega_L\hat a^{\dag}\hat a+i\hbar\kappa_L\mkern-2mu\left[\lambda(t)
e^{-i\omega_Lt}\hat a^{\dag}-\lambda^*(t)e^{i\omega_Lt}\hat a\right]\mkern-2mu,
\label{eqn:laser_Hamiltonian}
\end{eqnarray} 
where $\omega_L$ is the cavity resonant frequency and $\lambda(t)$ is the complex
amplitude of a time-dependent classical current driving the cavity mode. In a rotating
frame, with frequency $\omega_L$, it is readily shown (assuming the initial state to be
coherent) that the intracavity field is in the coherent state $|\alpha(t)\rangle_L$,
with $\alpha(t)$ satisfying the equation
\begin{eqnarray}
{\dot \alpha(t)}=\kappa_A\lambda(t)-\kappa_A\alpha(t).
\label{eqn:alpha_em}
\end{eqnarray}
For an initial vacuum state,
\begin{eqnarray}
\alpha(t)=\kappa_A\int_0^tdt'e^{\kappa_A(t'-t)}\lambda(t').
\label{eqn:alpha_sol}
\end{eqnarray}
We now return to master equation~(\ref{eqn:master}). Introducing the density
operator in the interaction picture,
\begin{equation}
\tilde\rho(t)=U_0^\dagger(t)\rho^\prime(t)U_0(t),
\end{equation}
where $U_0(t)\equiv\exp[-i(\omega_L\hat a^\dagger\hat a+\omega_A\hat\sigma_z/2)t]$,
we propose the {\it ansatz\/}
\begin{equation}
\tilde\rho(t)=|\alpha(t)\rangle\langle\alpha(t)|\otimes\rho_A(t),
\label{eqn:factored_state}
\end{equation}
which is suggested by our observation from quantum trajectories that, with the field
in a coherent state, the atom does not entangle with its driving field. Substituting
the {\it ansatz} into Eq.~(\ref{eqn:master}), and using Eq.~(\ref{eqn:alpha_em}) and the
relation
\begin{eqnarray}
\frac{d}{dt}|\alpha(t)\rangle=\left({\dot \alpha(t)}a^{\dag}
-\frac{1}{2}\frac{d}{dt}|\alpha(t)|^2\right)\mkern-2mu|\alpha(t)\rangle,
\end{eqnarray}
we obtain a master equation for the atom,
\begin{equation}
{\dot \rho_A}=\frac{1}{i\hbar}[H_{\rm eff},\rho_A]+{\cal L}_{\hat J_A}\rho_A,
\label{eqn:master4}
\end{equation}
with 
\begin{equation}
H_{\rm eff}=i\hbar\sqrt{4\kappa_L\kappa_A}\left[\alpha^*(t)\hat\sigma_-e^{-i\delta t}
-\alpha(t)\hat\sigma_+e^{i\delta t}\right],
\label{eqn:classical_H}
\end{equation}
where $\delta\equiv\omega_A-\omega_L$ and
\begin{equation}
\hat J_A=\sqrt\gamma\mkern2mu\hat\sigma_-.
\end{equation}
 
Master equation~(\ref{eqn:master4}), with effective Hamiltonian~(\ref{eqn:classical_H}),
is a central result. It demonstrates that the atom is effectively driven by a classical
field and that the total decoherence rate is determined by the spontaneous emission rate,
$\gamma=2\kappa_A+2\kappa'_A$, into all $4\pi$ modes. This is so despite the different
appearance of the decoherence for forwards scattering in Fig.~\ref{fig:fig2}, and despite
the $1/\bar n$ effect in the comparison between frames (a) and (b). We conclude that
consideration of the quantum nature of a coherent laser pulse does not uncover any
decoherence additional to what is accounted for by spontaneous emission.  

\section{Discussion}
\label{sec:discussion}
Our answer to the first of the two questions raised in Sec.~\ref{sec:background} has
just been given: the entanglement reported in \cite{Julio,Enk} does not give rise to
decoherence in addition to that due to spontaneous emission in the standard treatment
of the driving of an atom by a classical field. The second question from
Sec.~\ref{sec:background} touches on more subtle issues: it asks whether there is
entanglement between the atom and the laser field at all, or by extension, is any part
of the decoherence attributable to entanglement of the atom and the laser field?

Our comments below Eq.~(\ref{eqn:Fock}) and restated in the factored
state~(\ref{eqn:factored_state}) give a partial answer to this question. We find no
entanglement between the atom and the field that is in direct interaction with the
atom---the field at $z=l$ at time $t$ and in the cavity ($z=0$) at the retarded time
$t-l/c$. Such entanglement can arise for certain states of the driving field, but does
not arise for a coherent state. There remains the question of entanglement with the
field downstream from the atom, the forwards-scattered field. Is there entanglement
between the laser pulse and the atom after they interact?

In considering this question, we note first that at the level
of the unitarily evolving pure state of the entire system---laser, atom, and scattered
field (reservoir)---every correlation, no matter how prosaic, is accounted for through
entanglement. In the quantum trajectory description, entanglement between the atom and
scattered fields is resolved each $dt$---effectively continuously
in time as $dt\to0$---and the correlations accounted for through the label ${\rm REC}$ on
the system state. Quantum trajectories are therefore unable to say anything directly about 
entanglement between the forwards-scattered field and the atom. They do, however, retain the
correlations, and hence indicate the presence of entanglement indirectly; and certainly,
the atom and scattered fields are correlated, hence entangled at the level of the pure state
of the entire system. Even for spontaneous
emission (side scattering) the purpose of the quantum jump (\ref{eqn:jump_side}) is to
resolve entanglement at this level while retaining the corresponding correlation (for
one kind of measurement on the scattered field) between the scattering record and 
conditional state. For the forwards-scattered light, correlations are established in a
similar way; although in this case they take a very different form, showing, for example,
the mentioned $1/\bar n$ effect \cite{Car,Kochan}.
Specifically, the detection of a photon, which projects the atom into its ground state for
side scattering, can project the atom into its excited state for forwards scattering
\cite{Car}. For fixed Rabi frequency, the behavior is controlled---if indirectly---by
$1/\bar n$ (more directly, by the degree of focusing of the driving field onto the atom).
This difference in the form of the correlations comes from the overlap of the forwards
scattering with the incident field, which forms the post-interaction laser pulse. Thus,
even without an explicit expression for the state, it is safe to say that at the level
of the pure state for the entire system, the atom must be entangled with the post-interaction
laser pulse.

As a final comment we should emphasize that the analysis in this paper {\it presumes\/} that
the laser source is in a coherent state. The coherent state  is imposed by the model
Hamiltonian (\ref{eqn:laser_Hamiltonian}), not deduced. It has been suggested that an
operationally coherent laser is not strictly in a coherent state
\cite{Moelmer,Julio2,Moelmer1,Rudolf,Enk3,Sanders}. In principle, the COQS approach can
address this issue, by deducing the state of the source from a more realistic laser model.
While a deeper analysis along these lines might change what one has to say about the entanglement
structure of the pure state of the entire system, the central result that
there is no {\it additional\/} atomic decoherence is not expected to change.

This work was supported by the NSF under Grant No.\ PHY-0099576 and
by the Marsden Fund of the RSNZ.

\end{document}